\documentclass[11pt,a4paper]{article}

\usepackage[T1]{fontenc}
\usepackage[utf8]{inputenc}
\usepackage{lmodern}
\usepackage[margin=2.5cm]{geometry}
\usepackage{amsmath,amssymb}
\usepackage{booktabs}
\usepackage{graphicx}
\usepackage{xcolor}
\usepackage{listings}
\usepackage{tikz}
\usepackage[hidelinks]{hyperref}

\newcommand{\tetcube}[3]{%
  \draw[gray!60] (0,0,0) -- (1,0,0) -- (1,1,0) -- (1,1,1) -- (0,1,1) -- (0,0,1) -- cycle;
  \draw[gray!60] (1,0,0) -- (1,0,1) -- (0,0,1) (1,0,1) -- (1,1,1);
  \draw[gray!60, densely dashed] (0,0,0) -- (0,1,0) -- (1,1,0) (0,1,0) -- (0,1,1);
  \fill[tetfill, fill opacity=0.25] (0,0,0) -- (#1) -- (#2) -- cycle;
  \fill[tetfill, fill opacity=0.25] (0,0,0) -- (#1) -- (1,1,1) -- cycle;
  \fill[tetfill, fill opacity=0.25] (0,0,0) -- (#2) -- (1,1,1) -- cycle;
  \fill[tetfill, fill opacity=0.25] (#1) -- (#2) -- (1,1,1) -- cycle;
  \draw[tetedge!60, thin] (0,0,0) -- (#2) (#1) -- (1,1,1) (0,0,0) -- (1,1,1);
  \draw[tetedge, thick, -latex] (0,0,0) -- (#1) -- (#2) -- (1,1,1);
  \fill (0,0,0) circle (1.2pt) (1,1,1) circle (1.2pt);
  \node[below, font=\small] at (0.5,0,0) {\strut#3};
}
\colorlet{tetfill}{blue!70!black}
\colorlet{tetedge}{blue!60!black}

\lstdefinelanguage{Rust}{
  morekeywords={as,const,else,fn,for,if,impl,in,let,mut,pub,return,self,Self,struct,use},
  morekeywords=[2]{f32,u16,usize,Vec},
  sensitive=true,
  morecomment=[l]{//},
  morestring=[b]",
}
\title{Implementation Note 2: Approximating a Smooth Three-Dimensional Function on a Coarse Lattice}
\author{Dennis Luxen -- \href{mailto:dennis@project-osrm.org}{dennis@project-osrm.org}}
\date{18 September 2026}

\begin{document}
\maketitle

\section{Summary}

A smooth 3D function can be efficiently and accurately approximated by a 3D lookup table of significantly lower resolution combined with tetrahedral interpolation.
The function is sampled once on a coarse grid in a preprocessing step.
For each input, the enclosing grid cube is located, one of the 6 tetrahedra in this cube is selected, and the values at its 4 corners are combined by a weighted sum.

For RGB to CMYK color conversion, a particularly frequent application of this, a table with $n = 16$ cells per axis ($17^3$ nodes, 16-bit values) requires 38\,KiB.
A full 8-bit table requires 64\,MiB.
As shown in an experimental evaluation, the mean error of the $n = 16$ table is 0.08\,\% ink per channel.
For 8-bit input, the lookup uses integer arithmetic only (Section~\ref{sec:integer}).
It requires no floating point and no division, and all intermediate values fit in 32 bits.
Floating point is needed only to build the table, once and offline.
The integer result differs from tetrahedral interpolation with exact fractions by at most 1 unit of the 16-bit output, and white and black are reproduced exactly.
This makes the approach suitable for low-power devices and real-time applications where computational resources are limited.

\section{Problem}

A smooth function $f$ with 3 inputs and $k$ outputs is considered.
Evaluation of $f$ per pixel is too expensive, or $f$ is only available as measured data.
\[
  f : [0,1]^3 \to \mathbb{R}^k
\]
In color management, $f$ maps device RGB values to printer CMYK values, so $k = 4$.
$f$ is defined by an ICC profile or a printer model.
The color correctness of a laser printer\footnote{Ink and toner are used interchangeably.} is measured in the forward direction: test patches with known CMYK values are printed, and the resulting colors are measured.
$f$ requires the opposite direction, the CMYK value that produces a given color.
There is no closed-form inverse.
Each value of $f$ is found by an iterative search, and each iteration evaluates the forward model.
The search also has to select one of many CMYK values with the same color and handle colors outside the printer gamut (Section~\ref{sec:noformula}).
This is too slow for images with millions of pixels.
Likewise, storing a full-resolution 3D lookup table for all possible input values is impractical due to memory constraints.
For example, a full 8-bit table for RGB to CMYK would require $256^3 \cdot 4 \cdot 2$ bytes, which is 64\,MiB.
Typical flash memory sizes are half of this, making a full-resolution table impractical.

Objective: $f$ is stored at a small number of points.
All other values are reconstructed from these points at low cost and with low error.

\section{The lattice}\label{sec:lattice}

Each axis of the input space is divided into $n$ cells of width $h = 1/n$.
This gives $n + 1$ nodes per axis and $(n+1)^3$ nodes in total.
The last node on each axis is at 1.0.
Inputs on the boundary of the cube, for example $R = 255$, are therefore covered by stored values.
Each node stores $\Theta(k)$ output values, e.g., 4 for CMYK.
\[
  L[i][j][l] = f\left(\frac{i}{n}, \frac{j}{n}, \frac{l}{n}\right), \qquad i, j, l \in \{0, \dots, n\}
\]
The table is computed once, offline.
It contains $(n+1)^3 \cdot k$ values.
The amount of required memory grows cubically with $n$.
For example, the memory required for RGB to CMYK ($k = 4$) with 16 bits per output channel, i.e.\ 8 bytes per node:

\begin{center}
\begin{tabular}{lrr}
  \toprule
  $n$ (cells per axis) & Nodes & Size \\
  \midrule
  8 & $9^3 = 729$ & 5.7\,KiB \\
  16 & $17^3 = 4{,}913$ & 38\,KiB \\
  32 & $33^3 = 35{,}937$ & 281\,KiB \\
  255 (full 8-bit, 1 byte per value) & $256^3 = 16{,}777{,}216$ & 64\,MiB \\
  \bottomrule
\end{tabular}
\end{center}

The table stores 16-bit values for three reasons:
\begin{itemize}
  \item Error: an interpolated value is a weighted average of 4 stored values and inherits their rounding error.
  With 8-bit values, this error is up to $0.5/255 = 0.2$\,\% ink, more than twice the mean interpolation error at $n = 16$.
  With 16-bit values, it is at most 0.00076\,\% ink (Section~\ref{sec:errsources}).
  \item Compatibility: ICC profiles store their lookup tables with 8 or 16 bits per value, and 16 bits are common.
  \item Integer arithmetic: the method of Section~\ref{sec:integer} requires values below $2^{16}$, so that the weighted sum fits in 32 bits.
\end{itemize}

The full 8-bit table in the last row stores one entry for every possible input, so no interpolation takes place.
Each entry is the final output value, rounded once from the exact value.
For 8-bit output, 1 byte per value is therefore enough.
With 16-bit values, the full table would require 128\,MiB, which is even more impractical for typical flash memory sizes.

ICC profiles typically use 17 or 33 nodes per axis, i.e.\ $n = 16$ or $n = 32$.
Details of the integer arithmetic are given in Section~\ref{sec:integer}. The input is multiplied by $n$ and split into an integer part and a fraction.

\section{Tetrahedral interpolation}\label{sec:tetra}

\begin{figure}[!tb]
  \centering
  \begin{tikzpicture}[x={(1cm,0cm)}, y={(0.5cm,0.35cm)}, z={(0cm,1cm)}, line join=round]
    \begin{scope}[scale=2.2]
      \draw[gray!60] (0,0,0) -- (1,0,0) -- (1,1,0) -- (1,1,1) -- (0,1,1) -- (0,0,1) -- cycle;
      \draw[gray!60] (1,0,0) -- (1,0,1) -- (0,0,1) (1,0,1) -- (1,1,1);
      \draw[gray!60, densely dashed] (0,0,0) -- (0,1,0) -- (1,1,0) (0,1,0) -- (0,1,1);
      \draw[tetedge, thick] (0,0,0) -- (1,1,1);
      \foreach \r/\g/\b/\pos in {0/0/0/below, 1/0/0/below, 0/1/0/below right,
          1/1/0/right, 0/0/1/left, 1/0/1/above left, 0/1/1/above, 1/1/1/above right}
        \node[\pos, font=\footnotesize, inner sep=1.5pt] at (\r,\g,\b) {\r\g\b};
      \draw[-latex] (-0.75,0,0.25) -- node[below, font=\small] {$r$} (-0.35,0,0.25);
      \draw[-latex] (-0.75,0,0.25) -- node[left, font=\small] {$b$} (-0.75,0,0.65);
      \draw[-latex] (-0.75,0,0.25) -- (-0.75,0.45,0.25) node[above right, font=\small, inner sep=1pt] {$g$};
    \end{scope}
    \begin{scope}[xshift=4.2cm, yshift=2.1cm, scale=1.35]
      \begin{scope}[xshift=0cm]    \tetcube{1,0,0}{1,1,0}{$t_r \ge t_g \ge t_b$} \end{scope}
      \begin{scope}[xshift=2.3cm]  \tetcube{1,0,0}{1,0,1}{$t_r \ge t_b > t_g$}   \end{scope}
      \begin{scope}[xshift=4.6cm]  \tetcube{0,0,1}{1,0,1}{$t_b > t_r \ge t_g$}   \end{scope}
    \end{scope}
    \begin{scope}[xshift=4.2cm, yshift=-0.9cm, scale=1.35]
      \begin{scope}[xshift=0cm]    \tetcube{0,1,0}{1,1,0}{$t_g > t_r \ge t_b$}   \end{scope}
      \begin{scope}[xshift=2.3cm]  \tetcube{0,1,0}{0,1,1}{$t_g \ge t_b > t_r$}   \end{scope}
      \begin{scope}[xshift=4.6cm]  \tetcube{0,0,1}{0,1,1}{$t_b > t_g > t_r$}     \end{scope}
    \end{scope}
  \end{tikzpicture}
  \caption{Left: corner labels of a grid cube and the main diagonal from 000 to 111.
  Right: the 6 tetrahedra, in the order of the table, with the selection condition below each cube.
  The arrows show the path $000 \to P_1 \to P_2 \to 111$.
  Dashed edges are hidden behind the cube's foreground.}
  \label{fig:tetrahedra}
\end{figure}
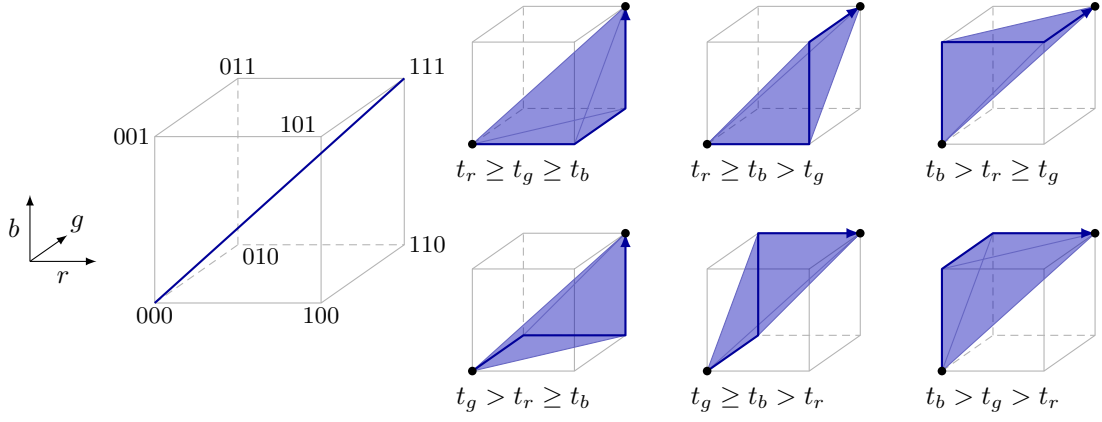

Each grid cube is divided into 6 tetrahedra.
All 6 tetrahedra share the main diagonal from corner 000 to corner 111 (Figure~\ref{fig:tetrahedra}).
The fractional position of the input inside the cube determines the tetrahedron as shown in the next Section.

\subsection{Step 1: cube index and fractions}\label{sec:step1}

The input is a point $(r, g, b) \in [0, 1]^3$, one coordinate per color channel.
For 8-bit RGB, $r = R/255$, $g = G/255$ and $b = B/255$.
Each channel is processed separately, along its own axis of the lattice.
The channel value is scaled to grid units.
The integer part is the cell index along the axis.
The remainder is the position inside the cell, in the range $[0, 1]$:
\[
  \begin{aligned}
    x_r &= r \cdot n, & i &= \min(\lfloor x_r \rfloor, n-1), & t_r &= x_r - i \\
    x_g &= g \cdot n, & j &= \min(\lfloor x_g \rfloor, n-1), & t_g &= x_g - j \\
    x_b &= b \cdot n, & l &= \min(\lfloor x_b \rfloor, n-1), & t_b &= x_b - l
  \end{aligned}
\]
Together, the three cell indices select the cube: its corner 000 is node $(i, j, l)$ and its corner 111 is node $(i+1, j+1, l+1)$.
The three fractions $(t_r, t_g, t_b)$ are the coordinates of the input inside this cube.
Steps 2 and 3 use only these fractions and the 8 corner values of the cube.

For $r = 1$, $x_r = n$, and due to the clamp to $n - 1$, the point is assigned to the last cell with $t_r = 1$, and node $n$ on the boundary is used.
The same applies to $g$ and $b$.

\subsection{Step 2: tetrahedron selection}\label{sec:tetsel}

The three fractions $(t_r, t_g, t_b)$ are sorted in descending order.
The order determines the tetrahedron selection.
Corner labels are offsets in $(r, g, b)$; for example, 101 denotes $(i+1, j, l+1)$.

\begin{center}
\begin{tabular}{ll}
  \toprule
  Condition & Corners on the path \\
  \midrule
  $t_r \ge t_g \ge t_b$ & $000 \to 100 \to 110 \to 111$ \\
  $t_r \ge t_b > t_g$   & $000 \to 100 \to 101 \to 111$ \\
  $t_b > t_r \ge t_g$   & $000 \to 001 \to 101 \to 111$ \\
  $t_g > t_r \ge t_b$   & $000 \to 010 \to 110 \to 111$ \\
  $t_g \ge t_b > t_r$   & $000 \to 010 \to 011 \to 111$ \\
  $t_b > t_g > t_r$     & $000 \to 001 \to 011 \to 111$ \\
  \bottomrule
\end{tabular}
\end{center}

In all 6 cases, the path starts at 000.
The first step is $+1$ along the axis with the largest fraction, the second along the axis with the middle fraction, and the third along the remaining axis.

\subsection{Step 3: weighted sum of the 4 corners}

The sorted fractions are denoted $t_1 \ge t_2 \ge t_3$, and the corners on the corresponding path $P_0, P_1, P_2, P_3$ along the tetrahedron are used.
The interpolated value is:
\[
  f(r,g,b) \approx (1 - t_1)\,P_0 + (t_1 - t_2)\,P_1 + (t_2 - t_3)\,P_2 + t_3\,P_3
\]
The 4 weights are the barycentric coordinates of the point in the tetrahedron.
All weights are non-negative, and their sum is 1.

\subsection{Properties}\label{sec:properties}

The tetrahedral interpolation has several important and beneficial properties:
\begin{itemize}
  \item Cost efficiency: 4 table reads and 3 multiplications per output channel.
  Trilinear interpolation, for example, requires 8 reads and 7 multiplications, which is more costly.
  \item Continuity: adjacent tetrahedra share faces.
  The interpolation result is continuous across cell and tetrahedron boundaries and avoids discontinuities within the tetrahedron.
  \item Linear precision: if $f$ is linear inside a tetrahedron, the result is exact.
  \item Neutral axis: for $r = g = b$, all fractions are equal, and only $P_0$ and $P_3$ have non-zero weights.
  In the color conversion example, both $P_0$ and $P_3$ lie on the gray diagonal.
  So,off-diagonal corners do not affect neutral colors.
\end{itemize}

\section{Worked example: RGB to CMYK Color Conversion}\label{sec:example}

For the input RGB (200, 120, 60) and $n = 16$ (17 nodes per axis), the result is CMYK (0, 40.06, 70.04, 21.57)\,\%.
The exact value is (0, 40.00, 70.00, 21.57)\,\%.
This example input is used throughout this and following sections for illustration.

\subsection{Excursion: why closed color conversion formulae don\textquotesingle t exist in practice}\label{sec:noformula}

RGB to CMYK for a real printer has no closed formula.
The transformation depends on measured printer behavior and on production settings.
The common formula $K = 1 - \max(R, G, B)$, $C = (1 - R - K)/(1 - K)$ describes an ideal printer with pure inks that combine linearly.
Real printers differ from this model in the following ways:

\begin{itemize}
  \item Ink interaction: overlapping inks do not absorb light independently.
  The color of an overprint depends on ink trapping, ink order and paper.
  The Neugebauer model~\cite{neugebauer1937} describes a halftone print by 16 primaries, one for each combination of C, M, Y and K, and each primary has to be measured.
  \item Dot gain: printed dots spread on the paper.
  The effective ink coverage differs from the nominal value.
  The difference depends on press, paper and screening.
  \item Inverse problem: a printer is characterized in the forward direction, from CMYK to a device-independent space such as CIELAB.
  A test chart is printed and measured, for example IT8.7/4 with 1,617 patches.
  RGB to CMYK requires the inverse of this measured mapping, which is computed numerically by an iterative search.
  \footnote{Quite often by repeatedly printing sheets of test patches with known CMYK values and measuring resulting colors under lab conditions.}
  \item Non-unique inverse: 4 outputs are computed from 3 inputs, so by pigeonhole principle many CMYK values produce the same color.
  The selected value depends on black generation (GCR or UCR) and on the total ink limit, for example 300\,\%.
  These are production settings, not physical laws.
  \item Gamut mapping: many RGB colors are outside the printer gamut.
  They are mapped to printable colors according to the rendering intent (perceptual, relative colorimetric, saturation).
  Perceptual mapping depends on the shape of the whole gamut and of course on the specific RGB color being mapped.
\end{itemize}

\subsection{Test function}\label{sec:testfunction}

Real printer profiles consist of measured data.
For verification against an exact reference, the following textbook formula for an ideal printer and ink is used as the test function $f$:
\[
  K = 1 - \max(R, G, B), \qquad C = \frac{1 - R - K}{1 - K}, \qquad M = \frac{1 - G - K}{1 - K}, \qquad Y = \frac{1 - B - K}{1 - K}
\]
For $R = G = B = 0$ (black), $1 - K = 0$, and the formulas for $C$, $M$ and $Y$ are undefined.
By definition, $C = M = Y = 0$ and $K = 1$: black is printed with black ink only.
The limit at black depends on the direction, so $f$ is discontinuous there (Section~\ref{sec:brightness}).
This formula is used only because its exact values are known, so the interpolation error can be measured.
For a real profile, the method is identical; only the table values differ.

\subsection{Step 1: fractions}

The input RGB color $RGB = (200, 120, 60)$ is normalized to $[0, 1]$ and multiplied by $n = 16$:

\begin{center}
\begin{tabular}{lrrll}
  \toprule
  Channel & Value / 255 & $\times 16$ & Cube index & Fraction \\
  \midrule
  R & 0.7843 & 12.549 & $i = 12$ & $t_r = 0.549$ \\
  G & 0.4706 & 7.529  & $j = 7$  & $t_g = 0.529$ \\
  B & 0.2353 & 3.765  & $l = 3$  & $t_b = 0.765$ \\
  \bottomrule
\end{tabular}
\end{center}

\subsection{Step 2: tetrahedron}

$t_b > t_r \ge t_g$.
The path is $000 \to 001 \to 101 \to 111$: first along B, then R, then G.

\subsection{Step 3: corners and weights}

\begin{center}
\small
\begin{tabular}{lllll}
  \toprule
  Corner & Node $(i, j, l)$ & Node RGB & Stored CMYK (\%) & Weight \\
  \midrule
  $P_0 = 000$ & $(12, 7, 3)$ & $(0.750, 0.438, 0.188)$ & $(0, 41.67, 75.00, 25.00)$ & $1 - 0.765 = 0.235$ \\
  $P_1 = 001$ & $(12, 7, 4)$ & $(0.750, 0.438, 0.250)$ & $(0, 41.67, 66.67, 25.00)$ & $0.765 - 0.549 = 0.216$ \\
  $P_2 = 101$ & $(13, 7, 4)$ & $(0.813, 0.438, 0.250)$ & $(0, 46.15, 69.23, 18.75)$ & $0.549 - 0.529 = 0.020$ \\
  $P_3 = 111$ & $(13, 8, 4)$ & $(0.813, 0.500, 0.250)$ & $(0, 38.46, 69.23, 18.75)$ & $0.529$ \\
  \bottomrule
\end{tabular}
\end{center}

\noindent The sum of the weights is 1.000.
Magenta: $0.235 \cdot 41.67 + 0.216 \cdot 41.67 + 0.020 \cdot 46.15 + 0.529 \cdot 38.46 = 40.06$\,\%.

\subsection{Result}

\begin{center}
\begin{tabular}{lrrrr}
  \toprule
  Method & C & M & Y & K \\
  \midrule
  Exact & 0 & 40.00 & 70.00 & 21.57 \\
  Tetrahedral (4 reads) & 0 & 40.06 & 70.04 & 21.57 \\
  Trilinear (8 reads) & 0 & 39.90 & 69.95 & 21.57 \\
  \bottomrule
\end{tabular}
\end{center}

The tetrahedral result deviates from the exact value by at most 0.06\,\% ink.
Trilinear interpolation achieves slightly worse accuracy andrequires twice as many table reads.

\section{Integer arithmetic for 8-bit input}\label{sec:integer}

Most printer hardware and software operate on 8-bit input values for each of the RGB channels.
Hardware is known to be low-cost and optimized for integer arithmetic rather than floating-point operations.
CPUs may not even have hardware support for floating-point operations and at times not even integer division.
This section describes how to perform the lookup using integer arithmetic for 8-bit input values using only basic arithmetic operations and no divisions.

The lookup is computed with integer arithmetic only retrieving stored 16-bit values.
The fractions are computed with 16 bits, in units of $1/65536$ of a cell.
The result is rounded once, at the end.

All computed quantities in this section are non-negative integers: the input values $R, G, B$, the table values, and every intermediate result.
Input channel refers to one of $R, G, B$, and output channel to one of the $k$ values stored per node.
$\lfloor a / 2^m \rfloor$ denotes a right shift by $m$ bits.
Quotients such as $T_r / 65536$ appear only in explanations and in the analysis, not in the computation.

\subsection{Step 1: cell index and fraction}\label{sec:step1int}

Step 1 is computed for each of the 3 input channels, separately and along its own axis.
For the input channel with the 8-bit value $R$:
\[
  y = nR, \qquad X = 257\,y + \left\lfloor \frac{y + 128}{256} \right\rfloor, \qquad
  i = \min\left(\left\lfloor \frac{X}{65536} \right\rfloor, n - 1\right), \qquad T_r = X - 65536\,i
\]
$j, T_g$ and $l, T_b$ are computed in the same way from $G$ and $B$.
$i$ is the cell index of Section~\ref{sec:step1}, and $T_r / 65536$ is the fraction $t_r$, rounded to 16 bits.
Each input channel requires 2 multiplications, 2 additions, 1 subtraction ($X - 65536\,i$), 1 comparison and 3 shifts, and no division.
$n - 1$ is a constant of the table.
$n$ must be a power of two and at most 128.

For $R = 200$ and $n = 16$: $y = 3200$, $X = 822400 + 13 = 822413$, $i = 12$ and $T_r = 35981$.
$T_r / 65536 = 0.54903$, and the exact fraction is $0.54902$ (Section~\ref{sec:example}).

\subsection{Steps 2 and 3: tetrahedron and weights}

Steps 2 and 3 use all 3 input channels together.
The tetrahedron is selected as in Section~\ref{sec:tetsel}, with $T_r, T_g, T_b$ in place of $t_r, t_g, t_b$.
The sorted fractions are denoted $T_1 \ge T_2 \ge T_3$.
The path starts at node $(i, j, l)$ and ends at node $(i+1, j+1, l+1)$, with one step of $+1$ per axis in the order of the sorted fractions.
$P_0, P_1, P_2, P_3$ denote the 16-bit values stored at these 4 nodes, for one output channel.
The weighted sum below is computed once per output channel.
The weights are integers:
\[
  W_0 = 65536 - T_1, \qquad W_1 = T_1 - T_2, \qquad W_2 = T_2 - T_3, \qquad W_3 = T_3
\]
The result is a 16-bit value for this output channel:
\[
  f(R, G, B) \approx \left\lfloor \frac{W_0 P_0 + W_1 P_1 + W_2 P_2 + W_3 P_3 + 32768}{65536} \right\rfloor
\]
The cell indices, the tetrahedron and the weights are the same for all output channels.
Only the 4 table reads and the weighted sum are repeated per output channel.
All intermediate values fit in unsigned 32-bit integers which is a fair assumption on minimum register and word sizes even for low-cost hardware and embedded devices.

Per input value $(R, G, B)$, the lookup therefore consists of 3 executions of Step 1, one tetrahedron selection, and $k$ weighted sums of 4 table values.
For RGB to CMYK, that is 3 cell indices with fractions, 1 tetrahedron, and 4 weighted sums.

\subsection{Correctness}

\begin{itemize}
  \item Position: the exact position in units of $1/65536$ of a cell is $65536\,nR/255$.
  Because $65536 = 257 \cdot 255 + 1$:
  \[
    \frac{65536\,y}{255} = 257\,y + \frac{y}{255}
  \]
  $257\,y$ is an integer, so rounding the exact position only requires rounding $y/255$.
  The formula for $X$ rounds $y/256$ instead.
  The two results differ only if a half-integer $k + \frac{1}{2}$ lies between $y/256$ and $y/255$, that is, for an integer $y$ with $255k + 128 \le y \le 256k + 127$.
  Because $y \le 255n$, this requires $k < n$.
  If $n$ is a power of two and at most 128, both $y$ and $256k + 128$ are multiples of $n$.
  Then $y \le 256k + 128 - n < 255k + 128$, so no such $y$ exists.
  $X$ is therefore the exactly rounded position for every input.
  An exhaustive check over all 256 input values and all powers of two from 1 to 128 confirms this.
  For other $n$, $X = \lfloor (65536\,nR + 127) / 255 \rfloor$ gives the exactly rounded position with an integer division by the constant 255.
  For $n \le 256$, $65536\,nR + 127$ is less than $2^{32}$ and fits the assumptions on 32-bit unsigned integer arithmetic.
  \item Boundary: $R = 0$ gives $X = 0$, which is node 0.
  $R = 255$ gives $y = 255n$ and $X = 65535n + n = 65536n$, which is node $n$.
  As in Section~\ref{sec:step1}, the clamp assigns this point to the last cell with $T_r = 65536$.
  If every channel is 0 or 255, one weight is 65536, and the stored node value is returned exactly.
  In particular, white and black are reproduced exactly, which is important for text documents.
  \item Weights and range: the fractions are integers from 0 to 65536.
  The weights are non-negative integers with sum exactly 65536.
  The weighted sum is a convex combination scaled by 65536, so it is at most $65536 \cdot 65535$.
  With the rounding constant 32768, it is less than $2^{32}$, again fitting the assumptions on word sizes.
  The result lies between the smallest and the largest of the 4 node values, and a constant table is reproduced exactly.
  \item Error: one LSB is one unit of the 16-bit output, that is $1/65535$ of the full scale, or 0.0015\,\% ink.
  $X$ is the exact position rounded to a multiple of $1/65536$ of a cell, so each fraction is off by at most $1/131072$ of a cell.
  Along each axis, the interpolated value is linear between two neighboring nodes, so a fraction error of $\delta$ cells changes the result by at most $\delta D$, where $D$ is the largest difference between neighboring node values.
  $D$ is at most the full range of 65535, which gives at most 0.5 LSB per axis.
  The final division adds at most 0.5 LSB.
  Over all $2^{24}$ inputs, for $n = 16$, the result differs from tetrahedral interpolation with exact fractions on the same table by at most 1 LSB (0.0015\,\% ink); the mean is 0.36 LSB.
  This is of the same order as the 16-bit quantization of the table as analyzed in detail in Section~\ref{sec:errsources} and far below the interpolation error.
  In practice, the differences are negligible and measurable only with precise instrumentation.
\end{itemize}

\subsection{Fraction width and position error}

With $F$ fraction bits, $X$ is the exact position rounded to a multiple of $2^{-F}$ of a cell, so the position is off by at most $2^{-F-1}$ of a cell.
This is $1/131072$ for $F = 16$ and $1/512$ for $F = 8$, a factor of 256.

With 8 fraction bits, the position is $X = y + \lfloor (y + 128)/256 \rfloor$ in units of $1/256$ of a cell, and the weights sum to 256.
This is exactly rounded by the same argument.
By the bound above, a position error of $1/512$ of a cell changes the result by up to $D/512$ per axis, which is 128 LSB for the largest possible $D$.
For $n = 16$ and inputs with a maximum channel value of at least 10\,\%, the measured change is up to 104 LSB and 9.4 LSB on average.
In ink, that is 0.16\,\% and 0.014\,\%, against a mean interpolation error of 0.08\,\% at $n = 16$.
16 fraction bits cost one multiplication by 257 more per input channel, or a shift and an addition, and are therefore used.

A common shortcut for $n = 16$ takes the high nibble of $R$ as the cell index and the low nibble as the fraction.
This is the position $X = 16R$ in units of $1/256$ of a cell, which scales the input by $n/256$ instead of $n/255$.
It is the exact position for an input range of 0 to 256, not for 0 to 255.
The exact position is $16R + 16R/255$, so the two differ by $16R/255$.
The position error therefore grows with $R$ and reaches 16 units, or $1/16$ of a cell.
For $R = 200$, the shortcut gives the fraction $8/16 = 0.5$ instead of $0.549$.
$R = 255$ gives cell 15 with the fraction $15/16$, so node 16 at 1.0 is never reached.
For white $(255, 255, 255)$, all three fractions are $15/16$, so the weight $1 - 15/16 = 1/16$ falls on node $(15, 15, 15)$ and the rest on node $(16, 16, 16)$.
For the test function, $K = 1/16$ at the first node and $K = 0$ at the second, so the result is $K = 1/256 = 0.39$\,\% instead of 0.
This is a visible tint on paper white.

The deviation is removed by adding $\lfloor (y + 128)/256 \rfloor$ with $y = 16R$, which is the rounded value of the difference $16R/255$.
The result is the position of Section~\ref{sec:step1int} with 8 fraction bits, and it is exactly rounded.
$R = 255$ then reaches node 16.
Compared with the 2 operations of the shortcut, a shift and a mask, the exactly rounded position costs 8 operations per input channel, and 9 with 16 fraction bits (Section~\ref{sec:step1int}).

\section{Implementation}

The Rust code below implements the method.
Function \texttt{build} computes the table (Section~\ref{sec:lattice}) in a preprocessing step.
\texttt{lookup} implements Section~\ref{sec:tetra} in floating point, and \texttt{lookup\_u8} implements Section~\ref{sec:integer} in integer arithmetic.
The idealized conversion function is \texttt{rgb\_to\_cmyk} for RGB to CMYK conversion.
The number of output channels \texttt{K} is a generic parameter, so the same code applies to RGB to CMYK ($K = 4$) and to other 3D functions.
In practice, the table is built once in an offline step on perhaps a differenent machine and then copied to the target device and used for all subsequent lookups.

\begin{lstlisting}[language=Rust]
/// A 3D lookup table on an (n+1)^3 lattice: n cells per axis,
/// n + 1 nodes per axis, K output values per node.
/// Values are stored as u16: 0 maps to 0.0 and 65535 maps to 1.0.
pub struct Lut3d<const K: usize> {
    n: usize,
    data: Vec<[u16; K]>,
}

const Q: f64 = 65535.0;

fn quantize(v: f64) -> u16 {
    (v.clamp(0.0, 1.0) * Q).round() as u16
}

impl<const K: usize> Lut3d<K> {
    /// Sample f at i/n, j/n, l/n for i, j, l in 0..=n (Section 3).
    /// Nodes at 1.0 are included, so edge inputs have stored values.
    pub fn build(n: usize, f: impl Fn([f64; 3]) -> [f64; K]) -> Self {
        assert!(n >= 1, "need at least 1 cell per axis");
        let h = 1.0 / n as f64;
        let mut data = Vec::with_capacity((n + 1).pow(3));
        for i in 0..=n {
            for j in 0..=n {
                for l in 0..=n {
                    let v = f([i as f64 * h, j as f64 * h, l as f64 * h]);
                    data.push(v.map(quantize));
                }
            }
        }
        Self { n, data }
    }

    /// Cells per axis.
    #[must_use]
    pub fn cells(&self) -> usize {
        self.n
    }

    /// Table size in bytes: (n + 1)^3 * K * 2.
    #[must_use]
    pub fn size_bytes(&self) -> usize {
        self.data.len() * std::mem::size_of::<[u16; K]>()
    }

    /// The K values stored at node c.
    #[must_use]
    pub fn node(&self, c: [usize; 3]) -> &[u16; K] {
        let m = self.n + 1; // nodes per axis
        &self.data[(c[0] * m + c[1]) * m + c[2]]
    }

    /// Tetrahedral interpolation at p in [0,1]^3 (Section 4).
    #[must_use]
    pub fn lookup(&self, p: [f32; 3]) -> [f32; K] {
        // Step 1: cell index and fractions per axis.
        let scale = self.n as f32;
        let mut base = [0usize; 3];
        let mut t = [0f32; 3];
        for (a, &pa) in p.iter().enumerate() {
            let x = pa.clamp(0.0, 1.0) * scale;
            // x == n on the edge: last cell with t = 1.
            let i = (x.floor() as usize).min(self.n - 1);
            base[a] = i;
            t[a] = x - i as f32;
        }

        // Step 2: axes sorted by fraction, largest first.
        let mut order = [0usize, 1, 2];
        order.sort_by(|&a, &b| t[b].total_cmp(&t[a]));
        let ts = [t[order[0]], t[order[1]], t[order[2]], 0.0];

        // Step 3: weighted sum along the path P0, P1, P2, P3.
        let mut corner = base;
        let mut out = [0f32; K];
        let w0 = 1.0 - ts[0];
        for (o, &v) in out.iter_mut().zip(self.node(corner)) {
            *o = w0 * f32::from(v);
        }
        for s in 0..3 {
            corner[order[s]] += 1;
            let w = ts[s] - ts[s + 1];
            for (o, &v) in out.iter_mut().zip(self.node(corner)) {
                *o += w * f32::from(v);
            }
        }
        out.map(|v| v / Q as f32)
    }

    /// Integer tetrahedral interpolation for 8-bit input (Section 6).
    /// Requires n to be a power of two, at most 128.
    /// Output on the table scale: 0 = 0.0, 65535 = 1.0.
    #[must_use]
    pub fn lookup_u8(&self, rgb: [u8; 3]) -> [u16; K] {
        debug_assert!(self.n.is_power_of_two() && self.n <= 128);
        // Step 1: position in units of 1/65536 cell, per axis.
        let n = self.n as u32;
        let mut base = [0usize; 3];
        let mut t = [0u32; 3];
        for (a, &r) in rgb.iter().enumerate() {
            let y = n * u32::from(r); // n * R
            let x = 257 * y + ((y + 128) >> 8); // round(65536 * n * R / 255)
                                                // R = 255 gives x = 65536 * n: last cell with t = 65536.
            let i = (x >> 16).min(n - 1);
            base[a] = i as usize;
            t[a] = x - (i << 16); // 0..=65536
        }

        // Step 2: axes sorted by fraction, largest first.
        let mut order = [0usize, 1, 2];
        order.sort_by(|&a, &b| t[b].cmp(&t[a]));
        let ts = [t[order[0]], t[order[1]], t[order[2]], 0];

        // Step 3: integer weights with sum 65536, then round.
        let mut corner = base;
        let mut acc = [0u32; K];
        let w0 = 65536 - ts[0];
        for (o, &v) in acc.iter_mut().zip(self.node(corner)) {
            *o = w0 * u32::from(v);
        }
        for s in 0..3 {
            corner[order[s]] += 1;
            let w = ts[s] - ts[s + 1];
            for (o, &v) in acc.iter_mut().zip(self.node(corner)) {
                *o += w * u32::from(v);
            }
        }
        // acc is at most 65536 * 65535, so acc + 32768 fits in u32.
        acc.map(|v| ((v + 32768) >> 16) as u16)
    }
}

/// Test function f (Section 5.2).
/// Inputs and outputs in [0,1].
#[must_use]
pub fn rgb_to_cmyk([r, g, b]: [f64; 3]) -> [f64; 4] {
    let k = 1.0 - r.max(g).max(b);
    if k >= 1.0 {
        return [0.0, 0.0, 0.0, 1.0]; // black: avoid 0/0
    }
    let d = 1.0 - k;
    [(1.0 - r - k) / d, (1.0 - g - k) / d, (1.0 - b - k) / d, k]
}
\end{lstlisting}

The demonstration program prints the values of the worked example:

\begin{lstlisting}[language=Rust]
use lut3d::{rgb_to_cmyk, Lut3d};

fn main() {
    let lut = Lut3d::build(16, rgb_to_cmyk); // 17^3 nodes
    println!("table size: {} bytes", lut.size_bytes());
    // expected: 39304
    let cmyk = lut.lookup([200.0 / 255.0, 120.0 / 255.0, 60.0 / 255.0]);
    println!("{:.2?}", cmyk.map(|v| v * 100.0));
    // expected: about [0.00, 40.06, 70.03, 21.57]
    // (16-bit rounding of the table can change the last digit)
    let cmyk = lut.lookup_u8([200, 120, 60]);
    println!("{cmyk:?}");
    // expected: [0, 26252, 45897, 14135], i.e. 40.06, 70.03, 21.57 %
}
\end{lstlisting}

\noindent\ The claims of Section~\ref{sec:integer} can be checked exhaustively, since the input domain has only $2^{24}$ points:

\begin{itemize}
  \item Position: for every 8-bit value $R$ and every power of two $n \le 128$, compare $X$ with $\lfloor (2 \cdot 65536\,nR + 255)/510 \rfloor$, which is $65536\,nR/255$ rounded to the nearest integer.
  \item Boundary: compare \texttt{lookup\_u8} at $(0,0,0)$ and $(255,255,255)$ with the values stored at the nodes $(0,0,0)$ and $(n,n,n)$.
  \item Error: compare \texttt{lookup\_u8} over all $2^{24}$ inputs with tetrahedral interpolation on the same table, computed in double precision with the exact fractions $t = nR/255 - i$, where $i$ is the clamped cell index of Section~\ref{sec:step1}.
  The largest deviation is 1 LSB.
\end{itemize}

For production use:
\begin{itemize}
  \item Use integer arithmetic for the per-pixel lookup. Floating point calculations are needed only to build the table.
  \item Replace the sort by the 6 comparisons from the case table.
  \item For a real printer, compute the table values from the ICC profile or with a color management library such as LittleCMS\footnote{\url{https://www.littlecms.com}}.
\end{itemize}

\section{Error analysis}

The interpolation error inside a tetrahedron is bounded by $\frac{3}{8} M h^2$, where $M$ is the largest curvature of $f$ in the cell.
The other error sources are either negligible or limited to the region near black.

\subsection{Error sources}\label{sec:errsources}

Three error sources are present in the method, and two more when the lookup is computed in integer arithmetic (Section~\ref{sec:integer}).

\begin{itemize}
  \item Interpolation: the curvature of $f$ inside a tetrahedron, analyzed below.
  \item Table quantization: the 16-bit node values add at most $0.5/65535 = 0.00076$\,\% ink, which is below 1\,\% of the mean interpolation error at $n = 16$.
  \item Test formula: the discontinuity at black (see Section~\ref{sec:brightness}).
  \item Fraction rounding, integer arithmetic only: the position is rounded to $1/65536$ of a cell, which adds at most 0.5 LSB per axis, for the largest possible difference between neighboring node values.
  \item Output rounding, integer arithmetic only: the final division adds at most 0.5 LSB.
\end{itemize}

One LSB is $1/65535$ of the output range, or 0.0015\,\% ink.
Over all $2^{24}$ inputs at $n = 16$, the two integer sources together change the result by at most 1 LSB, and by 0.36 LSB on average (Section~\ref{sec:integer}).
The interpolation error dominates: its mean at $n = 16$ is 0.08\,\% ink, more than 50 times the largest integer deviation.

\subsection{Theoretical bound}\label{sec:theory}

The error at a point $x$ inside a tetrahedron with vertices $v_i$ and barycentric weights $\lambda_i$ follows from a Taylor expansion of $f$ around $x$.
The linear terms cancel, because $\sum_i \lambda_i = 1$ and $\sum_i \lambda_i v_i = x$:
\[
  e(x) = f(x) - \sum_i \lambda_i f(v_i) = -\frac{1}{2} \sum_i \lambda_i \, (v_i - x)^T H(\xi_i) \, (v_i - x)
\]
$H$ is the Hessian of one output channel of $f$, evaluated at a point $\xi_i$ between $x$ and $v_i$.
With $M$ as the largest absolute eigenvalue of $H$ in the cell:
\begin{equation}\label{for:main_result}
  |e(x)| \le \frac{M}{2} \sum_i \lambda_i \lVert v_i - x \rVert^2 = \frac{M}{2} \left( R^2 - \lVert x - c \rVert^2 \right) \le \frac{3}{8} M h^2
\end{equation}
All 6 tetrahedra of a cube have the cube center $c$ as circumcenter, with circumradius $R = \frac{\sqrt{3}}{2} h$.
The bound is 0 at the nodes and largest at the cube center.
Sharp pointwise bounds for linear interpolation on a simplex are derived in \cite{waldron1998}.

Four results follow from the bound given by the main result in Equation~\eqref{for:main_result}:

\begin{itemize}
  \item Linear precision: if $f$ is linear in a cell, then $H = 0$, hence $M = 0$ and $e(x) = 0$.
  The interpolation reproduces such an $f$ exactly, at every point of the cell and for any $n$.
  A constant $f$ is reproduced exactly as well, because the weights sum to 1.
  The approximation therefore costs nothing where $f$ is linear, and the whole error is caused by curvature.
  \item Exactness on the lattice: the bound is 0 at the nodes, where the stored values are used directly.
  With $t_1 = t_2 = t_3$, only $P_0$ and $P_3$ have non-zero weights, so the interpolation is also exact along the main diagonal of each cube whenever $f$ is linear along it.
  For RGB to CMYK, this is the neutral axis (Section~\ref{sec:properties}).
  \item Second-order convergence: the bound is proportional to $h^2 = 1/n^2$.
  Halving the cell width reduces the error by a factor of 4 and increases the memory by a factor of 8.
  The measured slope of $-2.01$ confirms this (Section~\ref{sec:convergence}).
  \item Curvature as the only scale: the bound is proportional to $M$, the largest curvature of $f$ in the cell, and independent of the size of $f$ itself.
  A table is therefore as accurate as $f$ is smooth, and the error is concentrated where $f$ bends, which for the test function is the region near black (Section~\ref{sec:brightness}).
\end{itemize}

Linear precision is what makes the coarse lattice useful.
A real-worldprinter model is smooth over most of the color cube, so between the nodes $f$ is close to linear, and the remaining error is the second-order term of the bound.

\section{Experimental analysis}

For $n = 16$, the measured mean error is 0.08\,\% ink and the 99th-percentile error is 0.75\,\% ink, for inputs with a maximum channel value of at least 10\,\%.
The error decreases with $h^2$, as predicted by the bound.
Tetrahedral interpolation has less than half the mean error of trilinear interpolation.

\subsection{Setup}
The experimental setup to verify the quality claimsis summarized in the following table. The experiments were conducted on a quad-core Intel I7-4770 CPU, and are actually independent of the machine architecture.
\begin{center}
\begin{tabular}{lp{10cm}}
  \toprule
  Parameter & Value \\
  \midrule
  Test function & Textbook RGB to CMYK formula (Section~\ref{sec:testfunction}) \\
  Grid sizes & $n = 2, 4, 8, 16, 32, 64$ cells per axis \\
  Test points & 20,000 RGB points, uniform random in $[0, 1]^3$, fixed seed \\
  Error metric & Largest absolute error of the 4 channels per point, in \% ink \\
  Statistics & Mean and 99th percentile \\
  Excluded points & Maximum channel value below 10\,\%, except in Section~\ref{sec:brightness} \\
  Reference method & Trilinear interpolation on the same table \\
  Table precision & 64-bit floating point; the 16-bit quantization error is measured separately on 5,000 points \\
  Implementation & Rust\footnote{Compiler version rustc 1.98.1 (48a229cea 2026-09-01)}, with the node values in 64-bit floating point and the same interpolation as the listing \\
  \bottomrule
\end{tabular}
\end{center}

\subsection{Convergence}\label{sec:convergence}

\begin{figure}[p]
  \centering
  \begin{minipage}[t]{0.48\linewidth}
    \centering
    \includegraphics[width=\linewidth]{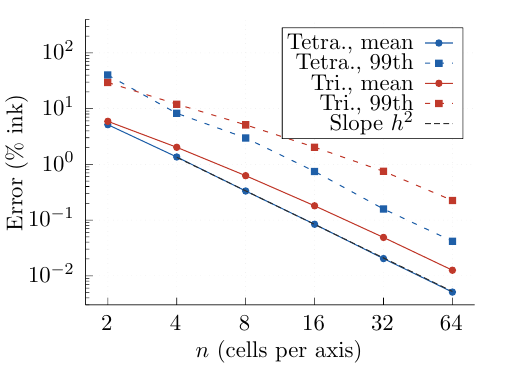}
    \caption{Error versus cells per axis, log-log.
    Worst channel per point; points with a maximum channel value below 10\,\% are excluded.
    The dark dashed line has the slope of $h^2$ and lies on the tetrahedral mean.}
    \label{fig:conv}
  \end{minipage}\hfill
  \begin{minipage}[t]{0.48\linewidth}
    \centering
    \includegraphics[width=\linewidth]{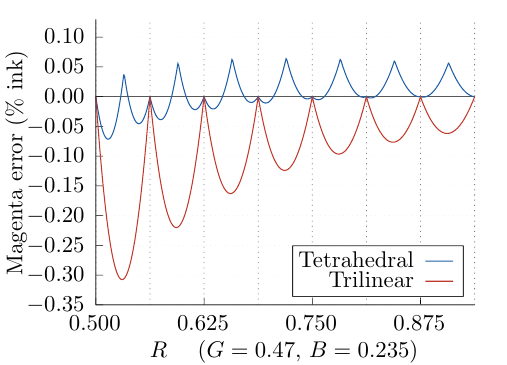}
    \caption{Signed magenta error along $R$ from 0.5 to 0.9375, at $G = 0.47$ and $B = 0.235$, $n = 16$.
    Dotted vertical lines mark the node planes $R = i/16$.}
    \label{fig:sweep}
  \end{minipage}

  \vspace{\floatsep}

  \begin{minipage}[t]{0.48\linewidth}
    \centering
    \includegraphics[width=\linewidth]{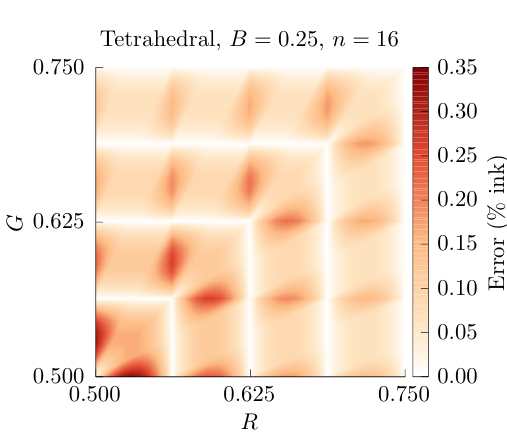}
    \caption{Worst-channel error of tetrahedral interpolation in the plane $B = 0.25$ (a node plane), $n = 16$, for $R$ and $G$ from 0.5 to 0.75 ($4 \times 4$ cells).
    White grid lines mark the cell boundaries.}
    \label{fig:errmap}
  \end{minipage}\hfill
  \begin{minipage}[t]{0.48\linewidth}
    \centering
    \includegraphics[width=\linewidth]{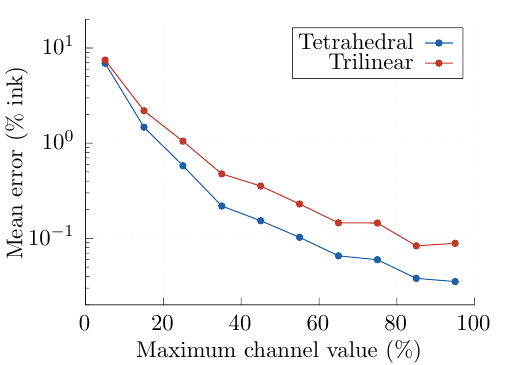}
    \caption{Mean worst-channel error per brightness bin (maximum of $R$, $G$, $B$), $n = 16$, all 20,000 points.
    Logarithmic axis.}
    \label{fig:bright}
  \end{minipage}
\end{figure}

Figure~\ref{fig:conv} shows the error for $n = 2$ to 64.
On the log-log axes, a slope of $-2$ corresponds to an error proportional to $h^2$.
The fitted slopes for $n = 8$ to 64 are $-2.01$ for the tetrahedral mean, $-2.07$ for the tetrahedral 99th percentile, $-1.88$ for the trilinear mean and $-1.50$ for the trilinear 99th percentile.

The slower decrease of the trilinear 99th percentile is consistent with the kinks of the test function.
$\max(R, G, B)$ switches on the planes $R = G$, $G = B$ and $R = B$, where $f$ is continuous but not differentiable.
These planes coincide with tetrahedron faces, so no tetrahedron contains a kink.
Trilinear interpolation combines values from both sides of a kink in the same cell.

\subsection{Error inside the cells}

Both methods are exact on the node planes (Figure~\ref{fig:sweep}) as predicted by the analysis of Section~\ref{sec:theory}.
For $R \ge G \ge B$, the magenta value $M = (R - G)/R$ is linear in $G$ and independent of $B$ for fixed $R$.
Between the planes, the trilinear error is negative, with a maximum of 0.31\,\% ink: $M$ is concave in $R$, and linear interpolation of a concave function underestimates it.
The tetrahedral error changes sign and stays below 0.08\,\% ink in all cases.
The maximum error for tetrahedral interpolation is therefore well below the trilinear maximum of 0.31\,\% ink, beating it by a factor of four.

Figure~\ref{fig:errmap} shows the plane $B = 0.25$ for $R$ and $G$ from 0.5 to 0.75.
The error is 0 where the largest channel is at a node value: on the lines $R = i/16$ for $R \ge G$, and on the lines $G = j/16$ for $G \ge R$.
Between these lines, the error increases towards lower values, because the curvature of $M = 1 - G/R$ and $C = 1 - R/G$ is larger for small $R$ and $G$.
The maximum in this area is 0.33\,\% ink at $(R, G) = (0.500, 0.531)$ and, by symmetry, at $(0.531, 0.500)$.

\subsection{Error by brightness}\label{sec:brightness}

The mean error in the darkest bin is about 200 times the mean error in the brightest bin (Figure~\ref{fig:bright}).
The test formula divides by $1 - K$, so C, M and Y are discontinuous at black.
A grid cannot represent a discontinuity.
Real profiles are continuous, but their curvature is also highest in the shadows.
Common measures are a finer grid, or a 1D curve per input channel before the lookup, which places more nodes in the dark range.
A practical solution could also be to apply a threshold beyond which the error is considered negligible, effectively ignoring the extreme values in the darkest regions.

\subsection{Choice of \texorpdfstring{$n$}{n}}

$n = 16$ is a suitable default.
$n = 32$ is used when the 99th-percentile error is relevant, for example for proofing.
Each doubling of $n$ reduces the error by a factor of 4 and increases memory by a factor of 8.
Above $n = 32$, the gain in accuracy is small compared with the memory cost.

\end{document}